\documentclass[letterpaper, a4paper]{amsart}

\usepackage{amsmath}
\usepackage{latexsym}
\usepackage[all]{xy}
\usepackage{color}

\include{amslatex}

\numberwithin{equation}{section}
\begin{document}
\begin{title}[Quantum illumination with multiple entangled photons]
 {Quantum illumination with multiple entangled photons}
\end{title}
\date{\today}
\maketitle
\thispagestyle{empty}
\begin{center}
\author{Ricardo Gallego Torrom\'e\footnote{Email: rigato39@gmail.com}}
\end{center}
\begin{center}
\address{Department of Mathematics\\
Faculty of Mathematics, Natural Sciences and Information Technologies\\
University of Primorska, Koper, Slovenia}
\end{center}
\bigskip
\begin{center}
\author{Nadya Ben Bekhti-Winkel\footnote{Email: nadya.ben.bekhti-winkel@fhr.fraunhofer.de}}
\end{center}
\begin{center}
\address{FHR, Fraunhofer-Institut f\"ur Hochfrequenzphysik und Radartechnik, Fraunhoferstr. 20, 53343 Wachtberg, Germany}
\end{center}
\begin{center}
\author{Peter Knott\footnote{Email: peter.knott@fhr.fraunhofer.de}}
\end{center}
\begin{center}
\address{FHR, Fraunhofer-Institut f\"ur Hochfrequenzphysik und Radartechnik, Fraunhoferstr. 20, 53343 Wachtberg, Germany}
\end{center}
\begin{center}
\address{Chair of Radar Systems Engineering
Institute of High Frequency Technology
RWTH Aachen University
Melatener Straße 25
52074 Aachen
Germany}
\end{center}

\begin{abstract}
In this work, a theoretical generalization of Lloyd's quantum illumination to signal beams described by two entangled photon states is developed. It is shown that the new protocol offers a method to find the range of the target, reduces the size of the required time-bandwidth product to have the same signal to noise ratio than in Lloyd's quantum illumination, has a lower probability of false positive and is resilient against noise and also potentially against losses. However, the generation of the required three photon states for the protocol posses a technical problem for its practical implementation not fully addressed. Recent advances in triple photon generation that can overcome this problem are discussed. Other issues related with the protocol are also considered.
\end{abstract}

\section{Introduction}
Lloyd's quantum illumination is a quantum sensing protocol that has opened new broad directions of research in quantum sensing and quantum communications \cite{Lloyd2008}. Quantum illumination is of special interest for radar applications, based upon the superiority in sensitivity that a quantum radar based upon quantum illumination has over any analogous {\it classical radar} based upon non-entangled sources of light with the same characteristics of intensity and frequency. Furthermore, since its inception, different quantum illumination protocols have been investigated and experimentally implemented \cite{Lopaeva et al.,Zhang et al. 2015}. Partially supported by such experiments and further theoretical developments  boosted the expectations for application of quantum illumination in quantum sensing and radar technology \cite{Marco Lanzagorta 2011}. Furthermore, the theoretical advantage of quantum illumination over classical equivalent illumination found potential application for quantum communication \cite{Zhuang et al. 2016}, making quantum illumination protocol a promising field by its applications in several areas of quantum technology.

Quantum illumination protocols rely on certain correlations between the idler and the signal states that survive the annihilation of entanglement due to noise and losses.
 For quantum sensing applications in the optical domain, the theory based upon Gaussian quantum illumination \cite{Tan} has been shown to be adequate. Tan et al. theory overcomes the a fundamental result of Shapiro and Lloyd on how coherent light can outperform Lloyd's quantum illumination when using light of the same energy and intensity characteristics \cite{ShapiroLloyd}. On the other hand, for radar applications, quantum illumination in the microwave regime \cite{Barzanjeh et al.} is the quantum illumination protocol that has been extensively investigated and demonstrations have illustrated a relative enhancement respect to classical illumination protocols  in laboratory conditions \cite{Barzanjeh et al.2019,Luong et al. 2019,Luong et al. 2020}.

 However, since its inception, fundamental limitations in the practical implementation of quantum illumination have been understood \cite{Pirandola et al. 2018,Shapiro2019,GallegoBenBekhtiKnott2020}. One of those fundamental problems is the  {\it range problem}: current applications of quantum illumination for target detection require beforehand knowledge of the target range, limiting the applicability of quantum illumination to scanning applications. Several interesting suggestions to overcome the range problem have been proposed. For instance, L. Maccone and C. Ren have suggested the use of beams composed by $N$-photon entangled states as signal beams without keeping any of the photons to form the idler beam \cite{MacconeRen2019}. Although Maccone-Ren protocol provides a theoretical method to obtain the range of the target and the transverse displacement of target from direct detection of the return state, the current form of their protocol is very sensitive to environmental noise and to losses. Partially entangled states have been suggested to solve the noise and loss problems, but such procedure reduces the advantage in sensitivity with respect to non-entangled light illumination.

We discuss in this paper a theoretical generalization of Lloyd's quantum illumination protocol where the signal beam is described by two photon states. The protocol that we present is a combination of  Lloyd's quantum illumination and Maccone-Ren's quantum radar protocol. Although coherent light outperforms in sensitive Lloyd's quantum illumination for light of the same characteristics \cite{ShapiroLloyd}, the attitude taken in this paper is to consider the simplest technical model of quantum illumination. This attitude is justified for the mathematical easiness of the model. Future developments of the theory can provide a full non-Gaussian squeezed states theory  formulation treatment that is required. Remarkably, an analogous procedure for signals described by $N>2$ entangled photon states follows along the same lines than the proposal discussed here. This is the reason why we have preferred to pay attention to the simplest case of $N=2$ signal photon states.

Our protocol can concisely be described as follows. A three entangled photon state is created by three generation through four photon non-linear optical interaction \cite{GarrisonChiao}. The three photon state is split  in an idler state, that contains one photon, and a signal state, that contains two  photons. Note that already by such splitting, the initial quantum entanglement of the triplet is lost but certain correlations of the quantum states remain and those are the ones responsible of the sensitivity enhancement. In particular, the three photons in each state are correlated in the time of generation (ideally), in frequency and in linear momentum. After preparation, the signal beam is sent to explore a given region of spacetime where a possible target can be located and the idler beam is retained in the laboratory. When the returned signal is detected, it is compared with the idler beam. In our criteria of detection, for a positive detection event, two photons arriving at the same time and being correlated in time and energy with the idler photon  must be detected. As a consequence, the protocol shows higher sensitivity in detection in terms of probability of false negative than Lloyd's quantum illumination.

As in the case of quantum illumination, the protocol presented in this paper is robust against environmental noise decoherence. It does not require that entanglement is preserved during the round trip detection process, since it only makes use of surviving correlations from the original three photon states as correlation in time and energy and certain correlations encoded in the form of the quantum states \cite{Weedbrook}. The protocol does not make specific use of the momentum correlation. Furthermore, the protocol provides a theoretical solution of the range problem in quantum radar based on the notion of radar distance. We also show that our protocol requires  shorter time-bandwidth product to have the same signal to noise ratio than in Lloyd's  quantum illumination protocol, in the domain of low signal to noise ratios and in situations where the number of modes of the signal is much larger than the average noise photon number, a natural condition naturally fulfilled in the optical regime. In the same domain, the probability of false positive is less than for Lloyd's quantum illumination. If the proposed protocol can be extended to situations of strong  environmental noise by means of a generalization of Gaussian quantum illumination theory \cite{Tan} to non-Gaussian illumination of the type associated to three (or multiple photon) squeezed states, then the proposed protocol will provide a theoretical solution for the time-bandwidth problem in quantum illumination, which is particularly acute issue in the microwave regime.

However, the protocol discussed in this paper is not free of fundamental and technical difficulties. Besides the threat of the idler losses, which is problem in common with other quantum illumination protocols, the generation of the  states for our quantum illumination protocol is the most important concern for the realization of our protocol. The detection method, that in our protocol is by direct coincidence detection, is one of the technical points that should be discussed further and eventually, maybe substitute by classical digital methods of detection that do not need to preserve the idler. From a theoretical point of view, we have formulated the theory in terms of direct photon detection, which has important technological constraints. These and other issues are considered in the discussion section.
\section{Quantum illumination with multiple entangled photons}\label{quantum illumination with multiple entangled photons}
 \subsection{Lloyd's quantum illumination using multiple entangled signal beams}
 We will consider the case when the background noise $N_B$ is small and the time-bandwidth product $M$ is large, as in the original analysis from Lloyd \cite{Lloyd2008}.
For the sake of technical considerations, the model that we consider is analogous to the original Lloyd's proposal instead than the description in terms of squeezed states \cite{Tan}. The signal states will describe two photon states. The generalization to signal beams described multiple photon states is straightforward.

We start considering a state of the form
\begin{align}
|\psi\rangle_3 =\,\frac{1}{\sqrt{M}}\,\sum^{M}_{\alpha=1}\,\lambda(\alpha)\,\hat{a}^\dag(\omega_1(\alpha),\vec{k}_1(\alpha))\,\hat{a}^\dag(\omega_2(\alpha),\vec{k}_2(\alpha))\,\hat{a}^\dag(\omega_3(\alpha),\vec{k}_3(\alpha))\,|0\rangle,
\label{three generation}
\end{align}
where $\alpha =1,...,M$ indicates the number of modes of the state and $\lambda(\alpha)\in\,\mathbb{C}$ such that
\begin{align*}
\sum^{M}_{\alpha=1}\,\lambda^*(\alpha)\,\lambda(\alpha)=\,M.
\end{align*}
The photons $1$, $2$, $3$ are correlated in time (they are generated at the same time), in frequencies by the relation $\omega_0=\,\omega_1(\alpha)+\omega_2(\alpha)+\omega_3(\alpha)$ and in momentum by the relation $\hbar\vec{k}_0=\,\hbar\vec{k_1}(\alpha)\,+\hbar\vec{k}_2(\alpha)\,+\hbar\vec{k}_3(\alpha) $  (see for instance, \cite{GarrisonChiao}, section 13.4), where $(\hbar\omega_o,\hbar\vec{k}_0)$ are the frequency and momentum of the pump beam.
After a convenient collimation, the momentum are selected in three fixed directions,
  \begin{align*}
  \vec{e}_1=\,\frac{\vec{k}_1}{\|\vec{k_1}\|},\vec{e}_2=\frac{\vec{k}_2}{\|\vec{k}_2\|},\vec{e}_3=\frac{\vec{k}_3}{\|\vec{k}_3\|},
  \end{align*}
   directions that are fixed for each value of the index $\alpha$.
The resulting state is transformed by an unitary operator $\mathcal{U}=\,I_1\otimes I_2\otimes U_3$ to prepare the sate in the form
\begin{align}
{|\Phi\rangle}_3 =\,\frac{1}{\sqrt{M}}\,\sum^{M}_{\alpha=1}\,\lambda(\alpha)\,\hat{a}^\dag(\omega_1(\alpha),\vec{k}_1(\alpha))\,\hat{a}^\dag(\omega_2(\alpha),\vec{k}_2(\alpha))\,\hat{a}^\dag(\omega_3(\alpha),\vec{k}'_3(\alpha))\,|0\rangle,
\label{three generation collimated}
\end{align}
where the photons $2$ and $3$ have parallel momenta, $\vec{k}_2 (\alpha)\, \| \,\vec{k}'_3 (\alpha)=\,\|\vec{k}_3(\alpha)\|\,\vec{e}_2$, but where the frequencies $\omega_2 (\alpha)$ and $\omega_3(\alpha)$ can be different. After this preparation,
the beam is split into two beams: the idler beam, which is composed by photons with $4$-momentum $(\hbar \omega_1,\,\hbar\vec{k}_1)$, and the signal beam, which is described by states with two photons with $4$-momentum $(\hbar\omega_2(\alpha)\,\hbar\vec{k}_2\|\vec{e}_2)(\alpha)$ and  $(\hbar\omega_3(\alpha),\hbar\|\vec{k}_3\|\vec{e}_2)(\alpha)$ respectively. The state \eqref{three generation collimated} must be considered equivalent to the state \eqref{idlersignal before quantum decoherence} in our appendix on Lloyd's theory.

After the state preparation, the system is describe by the density matrix of a pure state,
\begin{align*}
\rho_\Phi=\,{|\Phi\rangle}_3{\langle \Phi|_3}.
 \end{align*}
 Then the idler is retained in the laboratory system and the signal sent to explore a region in spacetime. One of the effects of noise environment is total decoherence on the state.
  Hence the idler system is well described approximately by the total mixed state
\begin{align}
\tilde{\rho}_1=\,\frac{1}{M}\,\sum^M_{\alpha =1}\,|\omega(\alpha),\vec{k}_1(\alpha)\rangle\langle \omega_1(\alpha),\vec{k}_1(\alpha)|,
\label{idler state}
\end{align}
while the state of idler-signal is denoted by $\tilde{\rho}^e$. In order to show enhancement, as in Lloyd's quantum illumination, it is not necessary to specify here the exact form of $\tilde{\rho}^e$.

Independently of the methodology followed for the preparation of the states to reach the state in the required form
\eqref{three generation collimated}, after the preparation and splitting of the beam, none of the photon states composing
 the idler beam and none of the states composing the signal beam  are entangled anymore. However, the following correlations persist  after the loss of entanglement:
\begin{itemize}
\item Time correlation in the generation of the three photons,

\item The photons on each state exhibit correlations in frequency,
\begin{align}
\omega_0=\,\omega_1(\alpha)+\,\omega_2(\alpha)+\,\omega_3(\alpha),\quad \alpha =1,...,M.
\label{correlation relation 1}
\end{align}

\end{itemize}
Note that although the photons were initially correlated in $3$-momenta, such correlation is lost during the preparation of the signal and idler beams.

Let us assume the above mechanism and methodology in the generation and preparation of the idler and signal beams.
The signal state is denoted by $\tilde{\rho}_2$, while the idler state is denoted by $\tilde{\rho}_1$. They can be both mixed states.
The noise environment is described by the state $\rho_0$ of the form
\begin{align*}
\rho_0  \approx \left\{(1-\,M_B\,N_B)|0\rangle\langle 0|+\,N_B\,\sum^{M_B}_{\beta=1}\,| a^\dag (\omega(\beta),\vec{k}(\beta))|0\rangle\,\langle 0|a(\omega(\beta),\vec{k}(\beta))|\right\},
\end{align*}
where the index $\beta =\,1,...,M_B$ indicates the possible noise spectra
\begin{align*}
Spect:=\{\omega(\beta)\in\, Noise\};
\end{align*}
 $M_B$ is the number of modes of the noise.
This is the noise quantum state used in Lloyd's theory (see the expression \eqref{state for noise} in Appendix \ref{AppendixLloyd quantum illumination}). Let us remark here that the signal beam  is collimated in such a way that the photons $2$ and $3$ are in the spectra of the noise,
\begin{align*}
\omega_2(\alpha),\,\omega_3(\alpha)\in \, Spect,\,\quad \alpha =1,...,M.
\end{align*}
Thus for practical purposes, one can take  the index $\alpha$ and the index $\beta$ as identical.

Given the structure of the signal beam and the underlying correlations with the idler beam that survive the preparation and the whole detection process, the criteria for a positive detection is the following,
\bigskip
\\
{\bf Criterion for positive detection.} {\it  We declare that the target is present if two photons in the spectrum range of the signal are detected back within an established time detection window at the same time than an idler photon is detected together in a joint measurement of the idler and signal beam and if the energy correlation relation \eqref{correlation relation 1} holds good.}
\bigskip

By joint measurement we mean here the following. When two photons are detected, one should compares such detection  with the possible detection of the corresponding idler photon. Ideally this second detection must happen at the same time than the measure of the detected signal pair. This procedure  requires to track when the idler photons are generated and for how long they stay in the laboratory an relate them by using correlation on time and energy with the pairs of photons detected. Note that direct method of detection could eventually be substituted by some other method that do not require to keep alive the beam during the whole process. An example of such methods are the {\it hybrid protocols} as developed in \cite{Barzanjeh et al.2019,Luong et al. 2019} and that use digital storage methods of the idler beam.

The notion of {\it time detection window}  must be adapted to the particularities of the given experimental situation. The time detection window cannot be too large to avoid confusion with uncorrelated photons. This constrains the operation characteristics of the direct photo detection device used and the available maximal range of detection. On the other hand, the time detection window cannot be too short as well, in order to avoid missing entangled sets of photons. Indeed, the generation process of three entangled photons imply a lower value for the time window of detection precision: if the generation time is within an interval $\delta t$ and the time detection window is $\Delta t$, then it is necessary that $\Delta t>\,\delta t$. Also, corrections due to the fact that the idler and signal beam propagate in  different media must be taken into account.

\subsection{Enhancement of the signal to noise ratio respect to classical light illumination}
In the following paragraphs we evaluate the signal to noise ratio for non-entangled illumination and for quantum illumination with multiple entangled photon states as signal. The procedure that we follow mimics the treatment in\cite{Lloyd2008}.
\\
{\bf A. Illumination with non-entangled light}. When the target is not  there and the illumination uses non-entangled light, the quantum states are described by the density matrix $\rho_0$. The probability of false positive  can be read directly from the structure of the state and, by the criteria of detection discussed above, it is associated with the event of detecting two photons in the same time window. Therefore, in the case of low bright environment ($N_B\ll\,1$), the probability of false positive is determined by the criteria of detection of two independent photons simultaneously with the required energies $\omega_2$, $\omega_3\,\in Spect$ and is given by the expression
\begin{align}
p_0(+)=\,(N_B)^2,
\label{N=3 non-entangled light nothere}
\end{align}
where it is assumed that the detection of each of the two photons are statistically independent events.

When the target is there, and under the same assumptions, the state is given by the density matrix
\begin{align*}
\rho_1& =\,(1-\eta)\rho_0+\eta\tilde{\rho}\\
& \approx (1-\eta)\left\{(1-\,M_B\,N_B)|0\rangle\langle 0|+N_B\sum^{M_B}_{\beta=1}| a^\dag (\omega(\beta),\vec{k}(\beta))|0\rangle\,\langle 0|a(\omega(\beta),\vec{k}(\beta))|\right\}+\eta\tilde{\rho} ,
\end{align*}
where $\tilde{\rho}$ stands for the state describing the signal when using non-entangled light.
The probability for the detection the arrival of two independent photons simultaneously is in this case of the form
\begin{align}
p_1(+)=\,((1-\eta)N_B+\,\eta)^2 .
\end{align}
The signal to noise ratio (SNR) in quantum illumination with two signal state photons when the illumination is performed with non-entangled light is given by the expression
\begin{align}
SNR_{CI2P}=\,\frac{p_1(+)}{p_0(+)}=\,\left(\frac{(1-\eta)N_B+\,\eta}{N_B}\right)^2 .
\label{SNR noentanglement in Maccone-Ren illumination}
\end{align}
One observes that this signal to noise ratio $SNR_{CI2P}$ is given by the square of the signal to noise ratio $SNR_{CI}$ in Lloyd's theory (expression \eqref{SNR noentanglement} in Appendix \ref{AppendixLloyd quantum illumination}). This is a natural consequence of the methodology followed. Therefore, when the signal to noise ratio is low in the sense that $SNR_{CI2P}<1$ there is a reduction in sensitivity respect to the original Lloyd's theory in the regime $SNR_{CI}<1$.
\\
{\bf B. Illumination with three photons entangled light}.
When there is no target present, the  noise-idler system is described by the density matrix $\tilde{\rho}^e_0$ of the form
\begin{align}
\tilde{\rho}^e_0 \approx \,\left\{(1-\,M_B\,N_B)|0\rangle\langle 0|+\,N_B\,\sum^{M_B}_{\beta=1}\,| a^\dag (\omega(\beta),\vec{k}(\beta))|0\rangle\,\langle 0|a(\omega(\beta),\vec{k}(\beta))|\right\}\otimes\,\tilde{\rho}_1,
\end{align}
where $\tilde{\rho}_1$ is the idler state given by the expression \eqref{idler state}

The probability of a false positive is the probability to attribute to the presence of the target the detection of two simultaneous returned photons. Within the scope of the approximations that we are considering, such a probability is independent of the details of the signal state and given by the expression
\begin{align}
p^e_0(+)=\,\left(\frac{N_B}{M}\right)^2.
\end{align}
This is just the square than in Lloyd's theory (equation \eqref{entanglemet0+}).

When the target is there, the state after decoherence and interaction signal-target is of the form
\begin{align*}
\tilde{\rho}^e_1 =\,(1-\eta)\cdot \tilde{\rho}^e_0+\,\eta\,\tilde{\rho}^e.
\end{align*}
By an analogous argument as in Lloyd's theory, the probability of detection using entangled light signal states when the target is there for one trial is
\begin{align}
p^e_1(+)=\,\left((1-\eta)\,\frac{N_B}{M}+\eta\right)^2.
\end{align}
Again, this is the square than the corresponding probability in LLoyd's theory (equation \eqref{entanglement1-}).
The signal to noise ratio is of the form
\begin{align}
SNR^e_{QI2R}=\,\frac{p^e_1(+)}{p^e_0(+)}=\,\left(\frac{M}{N_B}\,\right)^2\,\left((1-\eta)\frac{N_B}{M}+\eta\right)^2,
\label{multiple photon entangle illumination}
\end{align}
which is the square of the signal to noise ratio obtained for Lloyd's quantum illumination in the analogous case, equation \eqref{SNR entanglement}.

In the regime when the signal to noise ratio is low, expression \eqref{multiple photon entangle illumination} implies two different types of phenomena. The first is an enhancement from the use of quantum entangled states respect to non-entangled states, analogously as in Lloyd's theory. The second phenomenon is a reduction of enhancement respect to Lloyd's quantum illumination in the regime of low signal to noise ratio. However, the probabilities of false positive are lower for our protocol than for Lloyd's protocol.

In the regime of low signal to noise ratio, there is a reduction in the time bandwidth product needed to reach the same advantage in SNR than in Lloyd's quantum illumination. The time-bandwidth product is here the number of modes $M$ by quantum state. Let us compare the $SNR^e_{QI2R}$ of two entangle photon illumination \eqref{multiple photon entangle illumination} with the $SNR^e_{QI}$ in Lloyd's quantum illumination \eqref{SNR entanglement}. If both are to be of the same order, then the condition
\begin{align*}
SNR^e_{QI2R}=\left(\frac{M}{N_B}\,\right)^2\,\left((1-\eta)\frac{N_B}{M}+\eta\right)^2 = \,\frac{M'}{N_B}\,\left((1-\eta)\frac{N_B}{M'}+\eta\right)=\,SNR^e_{QI}
\end{align*}
where $M$ and $M'$ are a priori different, must hold good. 
If this condition is consistent with a reduction of the time band-width product in the sense that $M'>>M$, then
\begin{align*}
\left((1-\eta)+\eta\,\frac{M}{N_B}\right)^2 \gg \,\left((1-\eta)+\eta\,\frac{M}{N_B}\right).
\end{align*}
That leads, after simplifications leads to
\begin{align}
\frac{M}{N_B}\gg \,1.
\label{condition for comparison M M'}
\end{align}
The constraint \eqref{condition for comparison M M'} provides a sufficient condition for a reduction of the time bandwidth product, here associated to the number of modes $M$ and $M'$, such that  $SNR^e_{QI2R}=\,SNR^e_{QI}$ holds. Furthermore, the condition \eqref{condition for comparison M M'}  is easily satisfied in the optical regime, where time-bandwidth products of order $M\sim 10^{3}$ or larger are easily feasible.
 In the optical regime, typical values of $N_B$ at $\lambda =1.55\mu m$ is of order  \cite{Shapiro Guha Erkman 2005,Shapiro2019}
\begin{align*}
N_B=\,\pi\, 10^6\,\lambda^3\,N_\lambda/\hbar \omega^2\sim\,10^{-6}
\end{align*}
for daytime sky conditions, while in experimental demonstrations, $N_B\ll 1$ is a reasonable condition \cite{Lopaeva et al.,England Balaji Sussman 20019}.
In the microwave regime, it is also feasible to reach $M\sim 10^3$. However, for the microwave regime the  condition $N_B\ll\, 1$ is not satisfied in usual sky conditions. Therefore, in order to extend the protocol to the microwave regime, the condition $N_B\ll 1$ must be relaxed or drastically changed to $N_B\gg 1$.

\subsection{Determination of the range and transverse position using quantum illumination  with two photons signal states}
Let us assume that the scattering with the target returns two photons  such that the constraint \eqref{correlation relation 1} holds good and that the pair of returned photons, after scattering with the target, propagate in the coming back along the same direction. Then one can provide an straightforward method to determine the range of the target respect to the radar laboratory system at a given instant. Under these assumptions, the target range is given by the {\it radar distance},
\begin{align}
r_z=\, c\,\frac{1}{2}\left(T-t_0\right),
\label{range}
\end{align}
where $T$ is the arrival time of the two photons and $t_0$ is the time of emission. 

The procedure described above to determine the range of the target has important limitations. The first is that the photons in the signal state, after scattering with the target, can have different directions of propagation and do not necessarily are leaving the target in the same direction as they arrived, producing loses. The same problem appears also in the theory developed by Maccone and Ren \cite{MacconeRen2019}, where some of the photons in each quantum state can be lost after the scattering. In Maccone-Ren's theory, partial solutions of this problem are resolved by using nested systems of entangled photons, an idea that can also be adapted to our protocol.

\section{Discussion and conclusion}
The present work presents the first steps towards a theory of quantum illumination with signal beams described by quantum states with two entangled photons, although the fundamental idea and methods can be extended to the case of signal beams described by quantum states of $N>2 $ photons. In the regime of low intensity signal, low noise back-ground (but large compare with the noise signal) and low reflectivity of the target, we have shown that quantum illumination with beams described by two entangled photon states provides two relevant benefits respect to Lloyd's quantum illumination. The first benefit is a method to obtain the target range, based upon the notion of radar distance and target detection. This procedure can also be implemented in Lloyd's quantum radar protocols, as the experiments of England et al. discuss \cite{England Balaji Sussman 20019}. However, the methodology based upon detection of two signal photons instead of one provides a higher sensitivity.
 The second fundamental  benefit with respect to Lloyd's quantum illumination is the reduction of the  time-bandwidth product to have the same signal to noise ration than Lloyd's quantum illumination. This result is currently limited to the so-called {\it quantum illumination bad regime} subjected to the additional constraint \eqref{condition for comparison M M'}, that is, when the number of modes in the state is larger than the average number of modes of the noise state.

In order to effectively implement the protocol discussed in this paper for target detection systems, several problems remain to be addressed.
The first and most pressing difficulty is the generation of the required states \eqref{three generation collimated}. Three photons generation is an arduous task \cite{Bencheikh et al.,Borshchevskaya et al.}, but such states are of  fundamental interest in quantum technologies too, as for instance in  quantum computing \cite{Gottesman Kitaev Preskill} or quantum communication \cite{Hillery Buzek Berthiaume}. For quantum sensing, the difficulty strives in the generation of states with large time-bandwidth product with intensity large enough to be useful for quantum sensing. This difficulty is due to the small third order susceptibility $\chi^{(3)}$ of the associated non-linear process. Although the generation of three photon states by spontaneous down conversion has been demonstrated experimentally. In quantum illumination experiments where entangled pairs are generated by four wave mixing generation \cite{England Balaji Sussman 20019}, up to $97000$ per second coincident detection events idler/signal pairs for the idler/signal source, with the intensity strongly depending upon the pump power. Raman scattering conditioned the number of coincident pairs. Recent advances in triple generation suggests the possibility of bright triplets generation by than previous experiments and, by pumping with multiple modes, a multiple mode entangled states \cite{C.W. Sandbo Chang et al. 2020}. The brightness of the triplets obtained reaches up to 60 triplets per second per Hertz over a bandwidth up to hundred thousand Kilohertz, comparable to two photon down conversion generation \cite{Flurin et al.,C. W. Sandbo Chang et al. 2018}. The possibility of bright light three generation opens the possibility to be used as in the protocol described in this paper. Two further points must be met. The first is that the sources must be bright enough to allow detection after preparation and collimation. The second requirement is, as in LLoyd's quantum illumination, that the state should have large time-bandwidth product.

There are other methods to generate three photon states. Cascade down conversion generation of three photons can also be a source for the correlated states used in the protocol. This type of process has  been already observed \cite{Hubel et al. 2010}. The process consists of two consecutive spontaneous parametric down conversions to generate three photon final states. For the purposes of quantum illumination, cascade down conversion generation do not constrains on time of generation: only correlation in energy can be of utility. But if the intermediate energies are under control, two energy correlations need to be satisfied as follows. From one side, the first down conversion implies the correlation in frequencies $\omega_0 =\, \omega_1+\,\omega'_1 $; the second conversion implies $\omega'_1=\,\omega_2+\,\omega_3$. In practical terms, a protocol based upon  two energy correlation constraints  instead than time correlation detection is more difficult to implement for practical purposes.

Second, in order to apply quantum illumination with two photon states signal beams for long range radar applications, the signal beam must be in the microwave frequency regime. However, the original generated entangled state $|\psi_3\rangle$ does not need to be a three  microwave photons state, because existing current frequency conversion methods (electro-optomechanical converter \cite{Barzanjeh et al.}) can be applied from optical to microwave frequency. However, the application of such frequency conversion methods reduces the efficiency and the intensity of the beam.  Josephson parametric amplification (JPA) methods could provide an alternative to the laboratory generation of microwave entangled photons, but current JPA generation is un-capable to generate three photons entangled states.  Note that the present analysis of our protocol is restricted to the domain $N_B\ll\,1$, which is not a natural restriction for microwaves in atmospheric conditions. As remarked before, the unnatural restriction $N_B\ll\,1$ is the main motivation to extend our theory to consider squeezed states.

Third, as we mentioned before, if one of the of photons in a signal pair is lost, it makes the companion photon un-useful for detection according to our criteria for positive detection. In this sense, our protocol shares a similar problem  than  Maccone-Ren's quantum radar protocol. To solve the problem, Maccone and Ren suggested the use of nested entangled photon state systems  \cite{MacconeRen2019, MacconeRiccardi 2019}, that although reduces the performance, it is more robust protocol against losses. The same idea can potentially be adapted to our protocol. Even in the case that one photon of a pair of signal photons is lost, the arrival of a photon compatible with time and energy correlations with the idler beam can be used to declare a positive detection. In this case, the treatment well be a mixed of Lloyd's quantum illumination with one or two signal photon states. 

Fourth, there is considerable restriction due to the use of direct detection of photons. As discussed in \cite{GallegoBenBekhtiKnott2020}, direct detection limits the maximal range attainable. For instance, current avalanche photodiode detectors implies that the maximal range must be up to $300 \, m$. Of course, this cannot be seen as a fundamental limit of the detection method.

Besides the above specific problems and limitations related with the particularities of the protocol discussed in this paper, there are other issues in common with quantum illumination protocols. Among them,
the storage of the idler has been recognized as a dramatic constraint for quantum illumination protocols \cite{Barzanjeh et al.}. Either the use of radically improved quantum memories \cite{Barzanjeh et al.} or a combination of classical digital recording methods and matching filtering methods  as were introduced in \cite{Barzanjeh et al.2019,Luong et al. 2019, Luong et al. 2020} could provide a reasonable solution to the idler storage problem.

In conclusion, in the present work we discuss a new protocol for quantum radar that theoretically and under the above mentioned constraints, is resilient to noise and loses, requires less band-width product than quantum illumination, provides a methodology to determine the range of the target and is more sensitive (in probability of false positive) than Lloyd's quantum illumination. The theory put forward in the present work inevitably leads to further developments. One natural next step in the development of the theory is to consider the extension of the current model to the case when the environmental noise is large ($N_B\gg\,1$), in analogy of how the work of Tan et al. \cite{Tan} extended the protocol of quantum illumination to illumination with Gaussian states. In particular, we expect that by such extension, the protocol presented in this paper  can be extended to the microwave domain. On the other hand, the most urgent problems for the practical implementation of our protocol is to find a photon source mechanism for the required three photon states and the losses problem discussed above, a shared problem with other quantum technology developments.

\appendix\section{Sensitivity enhancement in Lloyd's quantum illumination: an illustrative example}\label{AppendixLloyd quantum illumination} In the following lines we discus in detail some aspects of Lloyd's theoretical protocol \cite{Lloyd2008}. We partially follow the exposition described in \cite{Marco Lanzagorta 2011}. As it is commonly used, hypothesis $0$ means that the target is not there, while when the target is there, the hypothesis is labeled by $1$.
\\
{\bf A. Non-entangled light illumination}. When the light used for illumination is described by non-entangled photons, the density matrix of the system idler-signal-noise, when the target is not there (hypothesis $0$) is
\begin{align}
\rho_0  \approx \left\{(1-\,M\,N_B)|0\rangle\langle 0|+\,N_B\,\sum^M_{n=1}\,|k\rangle_n\langle k |_n\right\},
\label{state for noise}
\end{align}
where $|k\rangle_n$ stands for a noise photon mode. Hence the probability of a false positive is
\begin{align}
p_0(+)=\,N_B,
\label{noentanglement0+}
\end{align}
 while the probability to be correct in the forecast that the target is not there is
\begin{align}
p_0(-)=1-\,p_0(+)=\,1-N_B.
\label{noentaglement0-}
\end{align}
If we repeat the experiment $m$ times, the probability of a false positive is
\begin{align*}
p_0(+,M)=\,(N_B)^m.
\end{align*}

If the target is there (hypothesis $1$), then the density matrix is given by
\begin{align}
\rho_1& =\,(1-\eta)\rho_0+\eta\tilde{\rho}\\
& \approx (1-\eta)\left\{(1-\,M N_B)|0\rangle\langle 0|+\,N_B\,\sum^M_{n=1}\,|k\rangle_n\langle k |_n\right\}+\,\eta \,|\psi\rangle_s\langle \psi|_s ,
\label{noentanglement 1 state}
\end{align}
where $|\psi\rangle_s$ stands for the state describing the signal, that one can assume first is a pure state, while $\eta$ is the reflective index. It follows that the probability to measure the arrival of photon is
\begin{align}
p_1(+)=\,(1-\eta)N_B+\,\eta
\label{noentanglement1+}
\end{align}
 and that consequently, the probability of false negative is
 \begin{align}
 p_1(-)=\,1-p_1(+)=1-((1-\eta)N_B+\,\eta)=(1-\eta)(1-N_B).
 \label{noentanglement1-}
 \end{align}
 The signal to noise ratio is given by the expression
 \begin{align}
SNR_{CI}=\,\frac{p_1(+)}{p_0(+)}=\,\frac{(1-\eta)N_B+\,\eta}{N_B}.
\label{SNR noentanglement}
\end{align}
\\
{\bf B. Entangled light illumination}. Let us now consider the case when the illumination is made using entangled states. For the case when there is no target there, the density matrix is given by the expression
\begin{align}
\rho^e_0 \approx \left\{(1-\,MN_B)|0\rangle\langle 0|+\,N_B\,\sum^M_{n=1}\,|k\rangle_n\langle k |_n\right\}\otimes\left(\frac{1}{M}\,\sum^M_{n=1}\,|k_n\rangle_I\langle k_n |_I \right),
\end{align}
where $\frac{1}{M}\,\sum^M_{k=1}\,|k\rangle_I\langle k |_I $ is the state of the idler.
The state
\begin{align*}
\rho_0=\,\left\{(1-\,MN_B)|0\rangle\langle 0|+\,N_B\,\sum^M_{n=1}\,|k\rangle_n\langle k |_n\right\}
\end{align*}
is the state describing the absence of the target. It determines the probability distributions to detect one photon due to noise only.
In Lloyd's theory, the initial prepared state is of the form
\begin{align}
\Psi_{IS}=\,\frac{1}{\sqrt{M}}\,\sum^M_{n=1}\,|k_n\rangle_I\otimes\,|k_n\rangle_{S}
\label{idlersignal before quantum decoherence}
\end{align}
The modes determining the idler/signal $k=1,...,M$ are selected to coincide with the modes of the noise. In this context, it is remarkable that the false positive probability for one individual detection,
\begin{align}
p^e_0(+)=\frac{N_B}{M}
\label{entanglemet0+}
\end{align}
 is dramatically reduced with the number of modes $M$. This was first highlighted by S. Lloyd in his seminal work \cite{Lloyd2008}.
 The probability of forecasting correctly the absence of the target  is given by the probability of the complement,
 \begin{align}
 p^e_0(-)=\,1-\frac{N_B}{M}.
 \label{entanglemet0-}
 \end{align}
Note than when the experiment is repeated a number $m$ of times in a independent way, the probability of a false positive  after detecting $m$ independent photons is
\begin{align*}
p^e_0(+,m)=\,\left(\frac{N_B}{M}\right)^m.
\end{align*}

When the target is there, for entangled states, the system idler-noise-signal is described by a density matrix of the form
\begin{align}
\rho^e_1 =\,(1-\eta)\cdot \rho^e_0+\,\eta\,\rho_s,
\end{align}
where $\rho_s$ is the density matrix of the signal photon system. Note that we do not specify here $\rho_s$, since the result of the calculation, interestingly, does not depend too much on them.
From this expression, one can extract the probability of detecting the target is
\begin{align}
p^e_1(+)=\,(1-\eta)\,\frac{N_B}{M}+\,\eta.
\label{entanglement1+}
\end{align}
The probability of no detection (interpreted as
a false negative) is of the form
\begin{align}
p^e_1 (-)=\,1-p^e_1(+)=\,(1-\frac{N_B}{M})\,(1-\eta).
\label{entanglement1-}
\end{align}
When applied $m$ independent experiments, the probability of right detection is
 \begin{align*}
p^e_1(+,m)=\,\left((1-\eta)\,\frac{N_B}{M}+\,\eta\right)^m.
\end{align*}
In the  case of false negative,
\begin{align*}
p^e_1 (-,m)=\,1-p^e_1(+)=\,(1-\frac{N_B}{M})^m\,(1-\eta)^m.
\end{align*}
The signal to noise ratio for this model of quantum illumination is
\begin{align}
SNR^e_{QI}=\,\frac{p^e_1(+)}{p^e_0(+)}=\,\left(\frac{M}{N_B}\right)\,\left((1-\eta)\frac{N_B}{M}+\eta\right),
\label{SNR entanglement}
\end{align}

From the above formulae and comparing the probabilities of false positive and detection using quantum enhancement respect to classical light, one observes a clear enhancement in sensitive when using quantum illumination \cite{Lloyd2008}.
Further details of the analysis of how the sensitivity enhancement arises using Lloyd's protocol can be found summarized in \cite{Lloyd2008}, in \cite{Marco Lanzagorta 2011}, section 5.5.3 and in \cite{GallegoBenBekhtiKnott2020}.

\subsection*{Acknowledgements}
We would like to thank L. Maccone, C. Ren and S. Pirandola for critical comments on the idea presented in this work. This work has been financed by {\it Fraunhofer Institute for High Frequency Physics and Radar Techniques FHR}.

\end{document}